\documentstyle[11pt,fewbody2005,twoside,epsfig]{article}

\def\ba{{\mathbf a}}
\def\bA{{\mathbf A}}
\def\bb{{\mathbf b}}
\def\bB{{\mathbf B}}
\def\bR{{\mathbf R}}
\def\ahat{{\hat\ba}}
\begin{document}

\title{Gravitational Scattering}

\author{Douglas C. Heggie}
\affil{Univ. of Edinburgh, Scotland}

\begin{abstract}
We review some modern applications of the theory of few-body encounters between binaries and
single stars.  In particular we focus on 
the treatment of adiabatic
encounters, in a regime which is of importance in encounters between a
star and a planetary system in a star cluster.
\end{abstract}
\section{Introduction: examples and applications}

Roughly speaking, gravitational scattering will be defined as the
study of few-body encounters in which particles interact by Newtonian
gravity, with certain types of initial conditions; namely, a few
(normally two or three) bound subsystems, such as single stars or
binaries, approach from infinity.  Then the problem is to characterise
the outcome, usually in a statistical sense. 

 The new book by Valtonen \&
Karttunen (2006) will very quickly become the standard reference on this
problem. 
While approximate analytical methods yield useful results in some
limiting situations (see, for example, Sec.\ref{sec:adiabatic}),
computer simulation is an essential tool.  And while efficient codes exist
for computing individual scattering encounters (e.g. {\sl triple}\footnote
{\tt http://www.ast.cam.ac.uk/$\sim$sverre/web/pages/nbody.htm}, {\sl
  fewbody}\footnote{\tt http://www.astro.northwestern.edu/$\sim$fregeau/code/fewbody/})
the scattering packages in {\sl starlab}\footnote{\tt
http://www.ids.ias.edu/$\sim$starlab/} provide the additional,
extremely valuable functionality of 
efficiently sampling parameter space.   Many possibilities exist for graphic
rendering of individual encounters, such as {\sl GLanim}\footnote{\tt
http://grape.astron.s.u-tokyo.ac.jp/$\sim$makino/softwares/GLanim/}. 

\subsection{Examples and applications} 

Though the topic was developed throughout the twentieth century, it remains topical, because of fresh applications, such as the following.

\subsubsection*{Scattering of
normal stars by a binary black hole} 

As a result of mergers, central binary black holes are expected in many
galaxies.  Observationally, binary black holes are studied at high
energies (e.g. NGC 6240, studied with Chandra: Komossa et al 2003) and at visual
wavelengths (e.g. the famous Tuorla object OJ287: Sillanpaa et al 1988).  A problem of
long standing is the evolution of the relative orbit of the black
holes, as they scatter stars from the surrounding galaxy (Hills 1983,
Gould 1991b, Mikkola \&
Valtonen 1992, Fukushige et al 1992, Quinlan 1996, 
Zier \& Biermann 2001, Merritt 2001, 2002, 
Yu \&
Tremaine 2003, Milosavljevi\'c \& Merritt 2003, Chatterjee et al 2003,
Gualandris et al 2005).

\subsubsection{Evolution of planetary systems in star clusters}
\label{subsec:clusters} For the solar system, though it is not now in a star
cluster, the question of stellar perturbations has been considered for
a long time (e.g. Lyttleton \& Yabushita 1965).  In recent years this question
has arisen because of the initially surprising absence of planets
(searched for photometrically) in star clusters (e.g. the globular
star cluster 47 Tuc; see Gilliland et al 2000, Weldrake et al 2005).  From the
theoretical point of view the study of such scattering is simplified
by the fact that one component of the participating binary is
effectively massless (Davies \& Sigurdsson 2001, Spurzem et al 2003,
Fregeau et al 2005).  We shall say more about this in
Sec.\ref{sec:adiabatic}  

\subsubsection{Capture of exotic particles by multiple systems} 
 The idea here is that the flux caused by interactions with normal matter may be
enhanced in situations where the particles can be trapped gravitationally, which allows them
multiple opportunities of interacting.  This is a more speculative
problem, but has still attracted considerable interest (Press \& Spergel 1985,
Gould 1987, 1988, 1991a, Damour \& Krauss 1998, 1999,  Gould \& Alam 2001, 
Lundberg \& Edsj\"o
2004).  Theoretically it is the case where
the incoming particle is nearly massless.  The complexity of the problem is
illustrated by a comparable problem of solar system dynamics: what fraction of
comets and asteroids are destroyed by colliding with the sun?  Several planetary
resonances are involved in the fact that almost half of a sample of
near-Earth asteroids
end by colliding with the sun (Farinella et al 1994).  It is also a common fate
of short-period comets (Levison \& Duncan 1994).

\subsubsection{The M4 triple}  The nearby globular cluster M4 contains a triple system in
which a distant companion, of mass comparable with or somewhat larger than
Jupiter's mass, is in orbit about a binary consisting of a white dwarf and a
millisecond pulsar.  It is likely to have formed in a four-body scattering
encounter between two binaries (Rasio et al 1995, Ford et al 2000, Fregeau et al
2005). 


\section{Adiabatic encounters}\label{sec:adiabatic}

\begin{figure} [h]
\begin{center}
\epsfig{file=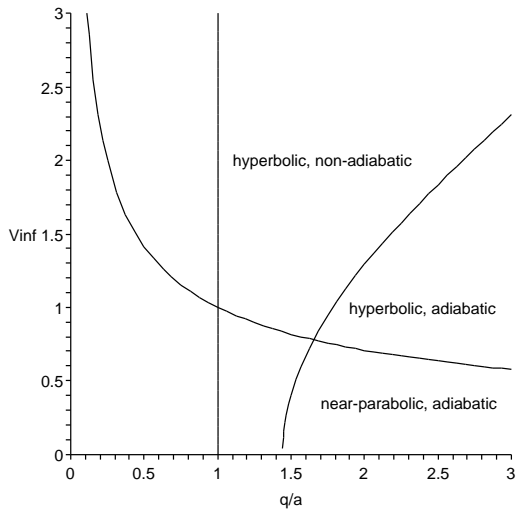,width=80mm,angle=0}
\caption {Regimes of tidal, adiabatic and hyperbolic encounters.  Encounters to
  the right of the vertical line are tidal.  These curves have been sketched for
  the case of equal masses, but do not depend sensitively on the masses unless
  $m_3$ is very large.  $V_{\inf}$ is plotted in units of
  $\sqrt{GM_{123}/a}$.}
\label{fig:regions}
\end{center}
\end{figure}

Consider a three-body scattering event involving a binary, with components of
mass $m_1, m_2$, and a third single star of mass $m_3$.  In this section we
shall be concerned with situations where the encounter is both {\sl tidal} and
{\sl adiabatic}.  Let the (initial) semi-major axis of the binary be $a$, and
$v$ the typical relative velocity of the components.  Then we shall estimate
$v^2\sim GM_{12}/a$, as in a circular orbit, where $M_{12} = m_1+m_2$.  Let the relative speed of the
star and the binary, when far apart, be $V_{\inf}$, and let $q$ be the distance of
closest approach between them.  Taking a Keplerian approximation for their
relative motion, we see that, at close approach, their relative speed is 
$V^2 = V_{\inf}^2 + \displaystyle{\frac{2GM_{123}}{q}}$, where
$M_{123} = m_1+m_2+m_3$, while the
eccentricity of their relative motion is $e' =
1+\displaystyle{\frac{qV_{\inf}^2}{GM_{123}}}$.  By plotting the curves $e'
  = 2$, $V/q = v/a$ (using the foregoing estimates) and $q=a$ we distinguish those
  encounters which are near-parabolic, adiabatic and tidal, respectively
  (Fig.\ref{fig:regions}).

Scattering problems like this can be approached analytically in various limiting
regimes.  High up in the diagram are non-adiabatic, impulsive encounters, which
covers most situations involving ``soft'' binaries.  At the bottom, just above the horizontal
axis, in the tidal regime, are near-parabolic, adiabatic encounters.  A
theoretical study of encounters in this regime was carried out by Roy \& Haddow
(2003), who gave explicit formulae for the change in energy of the binary during
the encounter.  Heggie \& Rasio (1996) had done a similar job for the change in
eccentricity of the binary, and had actually extended their result to the regime
of hyperbolic, adiabatic encounters.  This regime is important for one of the
problems mentioned earlier (Sec.\ref{subsec:clusters}), viz. the evolution of
planetary systems in star clusters, where typical values of $V_{\inf}$ are
comparable with the orbital velocity of a planet at a few AU.  The main purpose
of the present section is to extend the principal result of Roy \& Haddow to this regime.

Using first-order perturbation theory, and truncating the perturbing
potential between the binary and $m_3$ at quadrupole order, Roy \& Haddow show
that the change in energy of the binary is 
\begin{equation}
\delta\varepsilon = -
\frac{Gm_1m_2m_3}{M_{12}}\int_{-\infty}^{\infty}\dot\bR.\frac{\partial{\cal
R}}{\partial\bR} dt,\label{eq:de}
\end{equation}
where  $t$ is time, $\bR$ is the position vector of
$m_3$ relative to the barycentre of the binary, and $\cal R$ is the
perturbing function.  In turn, this can be taken to be of the form 
\begin{equation}\label{eq:perturbing_function}
{\cal R} = \frac{1}{R^5}\left\{\left[\frac{3}{2}e_1a^2(\hat\ba.\bR)^2 - \frac{3}{2}e_2b^2
(\hat\bb.\bR)^2 - \frac{1}{2}ee_3a^2R^2\right]\cos M +
3e_4ab\hat\ba.\bR\hat\bb.\bR\sin M\right\},
\end{equation}
where $a,b,e$ are, respectively, the semi-major axis, the semi-minor
axis and eccentricity of the binary, $\hat\ba,\hat\bb$ are unit
vectors along the axes of the orbit of the binary, $M$ is the mean
anomaly of the binary, and the remaining coefficients are defined to
be
\begin{eqnarray*}
e_1 &=& J_{-1}(e) - 2eJ_0(e) + 2eJ_2(e) - J_3(e)\\
e_2 &=& J_{-1}(e) - J_3(e)\\
e_3 &=& eJ_{-1}(e) - 2J_0(e) + 2J_2(e) - eJ_3(e)\\
e_4 &=& J_{-1}(e) - eJ_0(e) - eJ_2(e) + J_3(e),
\end{eqnarray*}
in which $J_n$ is the Bessel function of the first kind of order $n$.  

Roy \& Haddow proceed to use formulae for parabolic motion for $\bR$.
We follow their same procedure using formulae for hyperbolic motion
(e.g. Plummer 1918),
i.e.
\begin{eqnarray}\label{eq:bR}
\bR &=& a'(e'-\cosh F)\hat\bA + b'\sinh F\hat\bB\\
R &=& a'(e'\cosh F - 1),\label{eq:R}
\end{eqnarray}
where $\hat\bA, \hat\bB$ are unit vectors aligned with the axes of the
hyperbolic relative orbit of $m_3$, $a',b',e'$ are the hyperbolic
analogues of $a,b,e$ for this orbit, and $F$ is the eccentric anomaly,
related to time by
\begin{equation}\label{eq:kepler}
n't = e'\sinh F - F,
\end{equation}
in which 
\begin{equation}\label{eq:nprime}
n'^2a'^3 = GM_{123}
\end{equation}
and we have assumed $t=0$ at the
time of closest approach.  We also have $\cos M = \Re\exp(in(t-t_0))$,
where $t_0$ is the time of pericentric passage in the binary, and $n^2a^3=GM_{12}$.  

From eqs.(\ref{eq:de}) and (\ref{eq:perturbing_function}), we see that
a typical term in the integrand of $\delta\varepsilon$ is 
\begin{equation}\label{eq:typical_term}
\dot\bR\cdot\frac{\partial}{\partial\bR}\left(\frac{(\ahat\cdot\bR)^2}{R^5}\right)\cos
M = \Re\left(\frac{2\ahat\cdot\bR \ahat\cdot\bR^\ast R -
5(\ahat\cdot\bR)^2R^\ast}{R^6}\dot F\exp(in(t-t_0))\right),
\end{equation}
where $^\ast$ denotes differentiation with respect to $F$.  Indeed all the
terms in eq.(\ref{eq:de}) can be expressed in terms of the integral 
$$
{\cal I} = \int_{-\infty}^\infty\frac{f(F)\dot F}{R^6}\exp(in(t-t_0))dt,
$$
where $f$ is a polynomial in $\cosh F$ and $\sinh F$.  
By means of eq.(\ref{eq:kepler}), we then have
\begin{equation}\label{eq:F-eqn}
{\cal I} =
\exp(-int_0)\int_{-\infty}^\infty\frac{f(F)}{R^6}\exp\left\{
i\frac{n(e'\sinh F -F)}{n'}\right\}dF.
\end{equation}

\begin{figure} [h]
\begin{center}
\epsfig{file=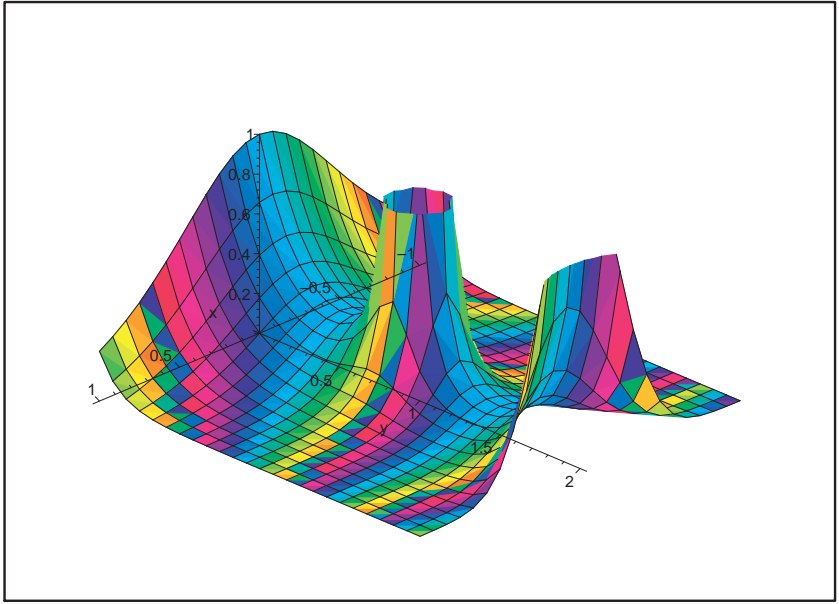,width=80mm,angle=-90}
\caption {The modulus of the integrand of eq.(\ref{eq:F-eqn}) in the
  region $\-1<\Re F<1, 0<\Im F<2$ 
  of the complex $F$-plane, for the case $e'\simeq1$.  Essentially the
  shape is that of a saddle, interrupted by a pole near $F=i$.}
\label{fig:complex}
\end{center}
\end{figure}

We handle this integral in the same way as in Heggie \& Rasio (1996).
Basically the method is steepest descents, but at the location of the
saddle of the exponent, it turns out that $R=0$.  (This is responsible
for the pole at about $F=i$ in Fig.(\ref{fig:complex}).)  First,
therefore, we clear off the factors of $R$ in the denominator.  We
deform the contour off the real $F$-axis to avoid the singularity at
$F=0$ in the following equivalent expression:
$$
{\cal I} =
\exp(-int_0)\int_{-\infty}^\infty\frac{f(F)R^\ast}{a'e'\sinh FR^6}\exp\left\{
i\frac{n(e'\sinh F -F)}{n'}\right\}dF,
$$
where we have used the derivative of eq.(\ref{eq:R}).  Integrating by
parts we can convert $R^\ast/R^6$ into $1/(5R^5)$, and in
differentiating the remainder of the integrand we can treat everything
except the exponential as constant:  the adiabatic assumption implies
that $n/n' \gg 1$, and so the derivative of the exponential factor
dominates.  Thus we have approximately
$$
{\cal I} =
\exp(-int_0)\frac{1}{5}\frac{in}{n'a'^2e'}\int_{-\infty}^\infty\frac{f(F)}{\sinh F}\frac{1}{R^4}\exp\left\{
i\frac{n(e'\sinh F -F)}{n'}\right\}dF,
$$
where we have made use of eq.(\ref{eq:R}) again. Doing this twice more
gives the approximate result
\begin{equation}\label{eq:after_parts}
{\cal I} =
\exp(-int_0)\frac{1}{1.3.5}\left(\frac{in}{n'a'^2e'}\right)^3\int_{-\infty}^\infty\frac{f(F)}{\sinh^3 F}\exp\left\{
i\frac{n(e'\sinh F -F)}{n'}\right\}dF.
\end{equation}

Now we deform the contour onto the saddle of the exponent, which
occurs where its derivative vanishes, i.e. $e'\cosh F - 1 = 0$.  We choose
the root $F_0 = i\arccos(1/e')$, which leads to 
\begin{equation}\label{eq:sinh_cosh}
\sinh F_0 = i\sqrt{1-\frac{1}{e'^2}},~~
\cosh F_0 = \frac{1}{e'}.
\end{equation}
A quadratic approximation of the
exponent around $F = F_0$ gives
$$
i\frac{n}{n'}(e'\sinh F -F) \simeq
-\frac{n}{n'}\left\{\sqrt{e'^2-1} - \arccos(1/e') +
\frac{1}{2}\sqrt{e'^2-1}(F-F_0)^2\right\}.
$$
Now the integral in eq.(\ref{eq:after_parts}) is easy, and gives
\begin{eqnarray*}
{\cal I} &=&
\exp(-int_0)\frac{1}{1.3.5}\left(\frac{in}{n'a'^2e'}\right)^3\sqrt{\frac{2\pi
n'}{n}}(e'^2-1)^{-1/4}\frac{f(F_0)}{\sinh^3
F_0}\times\\
&&\hskip1truein\times\exp\left(-\frac{n}{n'}(\sqrt{e'^2-1} - \arccos(1/e'))\right).
\end{eqnarray*}
Since $f(F)$ is defined implicitly in terms of $\bR, R$ and their
derivatives (cf. eq.(\ref{eq:typical_term})), it is helpful to know
that, when $F=F_0$, we have $\bR = a'(e'-1/e')(\hat\bA + i\hat\bB), R =
0, \bR^\ast = a'\sqrt{1-1/e'^2}(-i\hat\bA + \hat\bB)$ and $R^\ast =
ia'\sqrt{e'^2-1}$, where we have used eqs.(\ref{eq:bR}),(\ref{eq:R}) and (\ref{eq:sinh_cosh}), and the fact that $b' =
a'\sqrt{e'^2-1}$.  (The fact that $R=0$ at the saddle is the reason
for the three integrations by parts.)

Now we see that the integral (with respect to $t$) of
eq.(\ref{eq:typical_term}) is approximately
\begin{eqnarray*}
\Re\exp(-int_0)\frac{i\sqrt{2\pi}(e'^2-1)^{3/4}}{15a'^3e'^2}\left(\frac{n}{n'}\right)^{5/2}(-5)(\hat\ba.\hat\bA 
+ i\hat\ba.\hat\bB)^2\times\\
\times\exp\left(-\frac{n}{n'}(\sqrt{e'^2-1} - \arccos(1/e'))\right).
\end{eqnarray*}
We use eq.(\ref{eq:nprime}) and the corresponding relation for the
binary to eliminate $n$ and $n'$, and eliminate $a'$ in favour of the
distance of closest approach, $q$.  Thus the contribution of this term
to $\delta\varepsilon$ in eq.(\ref{eq:de}) is
\begin{eqnarray*}
-
 \frac{Gm_1m_2m_3M_{12}^{1/4}}{M_{123}^{5/4}}\frac{\sqrt{2\pi}}{10}\frac{(e'+1)^{3/4}}{e'^2}\frac{q^{3/4}}{a^{7/4}}(-5e_1)
\left\{((\hat\ba.\hat\bA)^2 
-(\hat\ba.\hat\bB)^2)\sin nt_0 -\right.\\
\left. - 2\hat\ba.\hat\bA\hat\ba.\hat\bB\cos nt_0\right\}\exp\left(-\left(\frac{M_{12}q^3}{M_{123}a^3}\right)^{1/2}\frac{\sqrt{e'^2-1} - \arccos(1/e')}{(e'-1)^{3/2}}\right).
\end{eqnarray*}
For ease of comparison, and a little economy, we introduce $K =
\displaystyle{\sqrt{\frac{2M_{12}q^3}{M_{123}a^3}}}$, and so this
contribution to $\delta\varepsilon$ is
\begin{eqnarray*}
-
 \frac{Gm_1m_2m_3}{M_{12}q^3}\frac{\sqrt{\pi}}{10}\frac{(e'+1)^{3/4}}{2^{3/4}e'^2}K^{5/2}(-5e_1a^2)\left\{((\hat\ba.\hat\bA)^2 
-(\hat\ba.\hat\bB)^2)\sin nt_0-\right.\\
\left. - 2\hat\ba.\hat\bA\hat\ba.\hat\bB\cos nt_0\right\}\exp\left(-\frac{K}{\sqrt{2}}\frac{\sqrt{e'^2-1} - \arccos(1/e')}{(e'-1)^{3/2}}\right).
\end{eqnarray*}
This is identical with the result of the corresponding terms in Roy \&
Haddow (the unnumbered equation below their eq.(17)), except for the
factor involving $e'$, and the exponent.  In the limit $e'\to1$ the
results are in agreement.

We can obtain a more convenient final expression by making the same
changes to eq.(19) in Roy \& Haddow, yielding
\begin{eqnarray}\label{eq:result}
\delta\varepsilon &=& - \frac{Gm_1m_2m_3}{M_{12}q^3}\frac{\sqrt{\pi}}{8}\frac{(e'+1)^{3/4}}{2^{3/4}e'^2}K^{5/2}
\exp\left(-\frac{K}{\sqrt{2}}\frac{\sqrt{e'^2-1} -
  \arccos(1/e')}{(e'-1)^{3/2}}\right)\times\nonumber\\
&&\times\left\{e_1a^2[\sin(2\omega+nt_0)(\cos2i-1)-
\sin(2\omega+nt_0)\cos2i\cos2\Omega -\right.\nonumber\\
&&-3\sin(2\omega+nt_0)\cos2\Omega -
4\sin2\Omega\cos(2\omega+nt_0)\cos i] + \nonumber\\
&&+e_2b^2[\sin(2\omega+nt_0)
(1-\cos2i)-\sin(2\omega+nt_0)\cos2i\cos2\Omega-\nonumber\\
&&-3\sin(2\omega+nt_0)
\cos2\Omega-4\cos(2\omega+nt_0)\sin2\Omega\cos i] +\nonumber\\
&&+
e_4ab[-2\cos2i\cos(2\omega+nt_0)\sin2\Omega -
6\cos(2\omega+nt_0)\sin2\Omega-\nonumber\\
&&\left.-8\cos2\Omega\sin(2\omega+nt_0)\cos
i]\right\},
\end{eqnarray}
where $\omega,\Omega$ and $i$ describe the orientation of the path of
$m_3$ in a frame aligned with the elliptical orbit of the binary.
Thus $\Omega$ is the longitude of the ascending node, measured in the
plane of the binary from its pericentre in the direction of its
motion.

It would be desirable to test this formula numerically.  So far the
only tests to have been carried out involve qualitative comparison of
scatter plots produced using both this formula and numerical
scattering experiments.  One reason why such a comparison is desirable
is the cavalier nature of the perturbation calculation we have carried
out.  In particular the end result is exponentially small in the ratio
$a/q$, whereas we
have neglected terms which are only algebraically small (e.g. higher
order terms omitted in eq.(\ref{eq:perturbing_function})).
Nevertheless the corresponding result for $e'=1$ works well (Roy \&
Haddow 2003).  Finally we remark that eq.(\ref{eq:result}) gives a
null result when $e = 0$.  Roy \& Haddow show how to deal with this
when the encounter is parabolic, and it is expected that a result for the general
case could be provided along the lines of the above calculation. 

\acknowledgments I warmly thank the organisers of the meeting for
making it possible to participate, and for their kind hospitality.  I
also thank Doug Lin, Rainer Spurzem and Mirek Giersz for attracting my
interest to the problem of stellar perturbations of a planetary
system.  Mirek in particular drew attention to the importance of the
eccentricity $e'$, and to an error in the original version of Fig.1.

\end{document}